\documentclass[pra,preprint,tightenlines,showpacs,twocolumn,10pt]{revtex4} 
\usepackage{epsfig} 

\begin{document} 
  
\title{ Four-body quantum dynamics of two-center electronic transitions 
in relativistic ion-atom collisions and target recoil momentum spectroscopy  } 
                                      
\author{ A.B.Voitkiv, B.Najjari, J.Ullrich } 
\affiliation{ Max-Planck-Institut f\"ur Kernphysik, 
Saupfercheckweg 1, D-69117 Heidelberg, Germany}  

\date{\today}  

\begin{abstract} 

We consider relativistic collisions of  
heavy hydrogen-like ions with hydrogen and helium 
atoms in which the ion-atom interaction causes both 
colliding particles to undergo transitions  
between their internal states. 
Using an approach enabling  
one, for the first time, to give a detailed description  
of this important case of the relativistic 
quantum few-body problem  
we concentrate on the study of 
the longitudinal momentum spectrum 
of the atomic recoil ions. We discuss the role 
of relativistic and higher order effects, 
predict a surprisingly strong influence 
of the projectile's electron 
on the momentum transfer, 
draw the general picture of 
the recoil ion formation  
and show that the important information 
about the doubly inelastic collisions 
could be obtained in experiment merely 
by measuring the recoil momentum spectrum.  

\end{abstract}


\maketitle 

% body of paper begins here 

The interaction between two point-like 
charged particles, whose relative velocity is 
much less than the speed of light $c=137$ a.u., 
is approximated with an excellent accuracy  
by the famous Coulomb law. 
As a result, this law underlies  
a vast amount of phenomena studied 
by atomic physics. 
Note, however, that even with 
such simplest form 
of the pair-wise interaction,  
one encounters large difficulties 
in attempts to get a detailed understanding  
of the dynamics of atomic collisions  
involving three and more 'active' particles 
\cite{nature}.  

When the relative velocity 
approaches the speed of light, 
the form of the pair-wise electromagnetic 
interaction becomes much  
more complicated. Consequently, a detailed 
description of relativistic atomic 
collisions involving three and more 
active particles, an important 
case of the fundamental quantum few-body problem 
(in its relativistic dynamic version), 
represents a particularly 
strong challenge for theory and 
many phenomena occurring in such collisions 
are much less understood compared 
to their non-relativistic counterparts.  

Highly charged projectiles produced 
at accelerators of heavy ions often 
carry one or more very tightly bound electrons. 
When such a projectile-ion impinges on 
a target-atom the interaction between 
them can cause both the ion and atom 
to undergo transitions between 
their internal electronic states. 
Such ion-atom collisions 
are termed 'doubly inelastic'. 

Doubly inelastic collisions of light ions 
with atoms at nonrelativistic impact energies 
have been extensively studied 
experimentally and theoretically  
(see e.g. \cite{MMM} and references therein). 
However, for the relativistic domain of 
impact energies, especially for 
collisions involving very highly charged ions,  
not only no results of coincidence measurements 
for doubly inelastic processes have been 
reported so far but also the existing 
theoretical approaches actually did 
not enable one to address them. 

From the theoretical point of view 
the main difficulties are connected 
with the complicated form of 
the pair-wise relativistic interaction 
and also with the fact that, at any values of 
the relative ion-atom velocity, doubly 
inelastic processes in collisions with   
very highly charged ions cannot be correctly 
described within first order approaches in 
the projectile-target interaction.  
 
The main obstacle for detailed experimental 
studies of the doubly inelastic processes 
is that cross sections 
for excitation and loss of an electron, 
tightly bound in a highly charged projectile, 
are by orders of magnitude smaller than those for 
ionization of a neutral atom. 
Therefore, it might seem to be extremely difficult 
to extract a small amount of doubly inelastic events 
out of a very large background produced 
by singly inelastic collisions in which 
only the atom undergoes electronic transitions.  

In this letter we consider    
doubly inelastic processes in collisions between 
a very heavy hydrogen-like ion,    
whose nucleus has a charge $Z_I \gg 1$ a.u. 
and moves with a velocity approaching 
the speed of light, and hydrogen and 
helium atoms. In particular, it will be shown 
that, due to the surprisingly 'prominent'  
role of the projectile electron 
in the momentum transfer, 
the important information about  
these processes could be obtained 
in an experiment just by measuring the 
longitudinal momentum spectrum of 
the target recoil ions.  

In high-energy atomic collisions 
the changes in the velocities of 
the atomic and ionic nuclei are negligible. 
Therefore, in order to describe 
these collisions it is convenient to introduce  
the following two reference frames. 
In one of them, denoted by $K_A$, 
the nucleus of the atom 
having a charge $Z_A$ 
is at rest and taken as 
the origin of $K_A$.  
In the other frame, $K_I$, 
the nucleus of the ion rests and its position 
is taken to be the origin of $K_I$.   

Let $\varphi_0$ and $\varphi_n$ ($n \neq 0$) 
represent the initial and final  
internal states of the atom in 
the frame $K_A$ whose energies  
in this frame are $\varepsilon_0$ and 
$\varepsilon_n$, respectively. 
Let $\chi_0$ and $\chi_m$ ($m \neq 0$)  
be the initial and final internal states 
of the ion in the frame $K_I$ and $\epsilon_0$ and 
$\epsilon_m$ be their energies in $K_I$. 
In the frame $K_A$ the incident projectile has a velocity 
$ {\bf v} = (0,0,v) $, and $ \gamma =1/\sqrt{1-v^2/c^2} $ 
is the corresponding collisional Lorentz factor.  
 
At $Z_I/v \ll 1$ (atomic units are used) 
the physics of doubly inelastic  
ion-atom collisions is mainly governed 
by the so called two-center 
dielectronic interaction, which  
couples two electrons 
orbiting the different colliding centers  
and occurs predominantly via a single 
photon exchange \cite{VNU}, 
and can be well described within the 
first order approaches  \cite{Voit-review1}. 
However, since $v < c$, the condition $Z_I/v \ll 1$ 
is never fulfilled in collisions with 
very highly charged projectiles for which 
a much more sophisticated treatment is needed. 

By using a refined version 
(to be discussed in detail elsewhere) 
of the relativistic eikonal model, 
developed recently in \cite{abv-eik}, 
one can show that the transition amplitude for 
the doubly inelastic collisions 
between a hydrogen-like ion and 
a two-electron atom is given by  
\begin{eqnarray}
S_{fi}({\bf Q}) %
= && - \frac{2i}{v} %
\int d^2 {\bf p}_1 %
\int d^2 {\bf p}_2  %  
\int d^2 \mbox{\boldmath$\kappa$} %  
\nonumber \\  
&& \frac{ f(p_1,\nu) \, f(p_2,\nu) % 
\, f(\kappa,-\eta) } % 
{ ({\bf q}^A + \mbox{\boldmath$\kappa$} % 
- {\bf p}_1 - {\bf p}_2)^2 % 
- \frac{(\varepsilon_n-\varepsilon_0)^2}{c^2} } % 
\nonumber \\ 
&& \times \Phi^A_{\mu}\left( n0; % 
{\bf q}^A+ \mbox{\boldmath$\kappa$} % 
-{\bf p}_1- {\bf p}_2; % 
{\bf p}_1; {\bf p}_2 \right) %
\nonumber \\ 
&& \gamma^{-1} \, \Lambda^{\mu}_{\,\alpha} \, % 
F_I^{ \alpha }% 
\left( m0; {\bf q}^I - \mbox{\boldmath$\kappa$} % 
+ {\bf p}_1 + {\bf p}_2) \right),   
\label{e1} 
\end{eqnarray}
where ${\bf Q}$ is the transverse part,  
${\bf Q} \cdot {\bf v}=0$, 
of the total momentum ${\bf q}^A$ 
transferred to the atomic target in the collision.  
In (\ref{e1}) the integration runs over 
the two-dimensional transverse vectors 
${\bf p}_1$, ${\bf p}_2$ and 
$ \mbox{\boldmath$\kappa$} $ 
($0 \leq p_j < \infty$, ${\bf p}_j \cdot {\bf v} =0 $;  
$0 \leq \kappa < \infty$, 
$ \mbox{\boldmath$\kappa$} \cdot {\bf v} =0$),     
$\nu = Z_I/v$, $\eta=Z_I Z_A/v$ and     
\begin{eqnarray} 
f(a,\tau) = 
\frac{ \Gamma( 1 - i \tau ) \Gamma(1/2+i \tau)}%
{2 \pi \Gamma(1/2) \Gamma( 2 i \tau ) } %
a^{\delta - 2 + 2i\tau}, 
\label{e2}
\end{eqnarray} 
where $ \delta \to +0 $ and 
$\Gamma(x)$ is the gamma-function.  
Further, $F_I^{ \alpha }$ and $\Phi^A_{\mu}$ 
($\alpha,\mu=0,1,2,3$) are the inelastic form-factors 
of the ion and atom, respectively, 
$\Lambda^{\mu}_{\,\alpha}$ is 
the Lorentz transformation matrix 
and the summation over the repeated 
Greek indices is implied in Eq.(\ref{e1}). 
The explicit form of the coupling 
of the form-factors in Eq.(\ref{e1}) 
is somewhat cumbersome  
\begin{eqnarray}
\Phi^A_{\mu} \, \gamma^{-1} \, % 
\Lambda^{\mu}_{\alpha} \, F_I^{\alpha} &=& %
\left(\Phi^A_0 + \frac{v}{c} \Phi^A_3 \right) %
\left(F^0_I + \frac{v}{c} F^3_I \right) +%
\nonumber \\ 
&&  \frac {\Phi^A_3 F^3_I}{\gamma^2} + %
\frac {\Phi^A_1 F^1_I + \Phi^A_2 F^2_I }{\gamma}   
\label{e3}
\end{eqnarray} 
reflecting rather complicated 
nature of the interaction between two 
relativistic transition four-currents. 

The components of the atomic two-electron 
form-factors are given by   
\begin{eqnarray} 
\Phi^A_{0}(n0;{\bf k};%  
{\bf p}_1;{\bf p}_2) %
= && \langle \varphi_n \mid  %
Z_A \exp( i {\bf p}_1 \cdot {\bf r}_1 + %  
i {\bf p}_2 \cdot {\bf r}_2) % 
\nonumber \\ 
&&- \exp( i {\bf p}_1 \cdot {\bf r}_1 % 
+ i {\bf p}_2 \cdot {\bf r}_2 ) \times % 
\nonumber \\ 
&& \left( \exp( i {\bf k} \cdot {\bf r}_1)  
+ \exp( i {\bf k} \cdot {\bf r}_2) \right) % 
\mid \varphi_0 \rangle 
\nonumber \\ 
\Phi^A_{l}(n0;{\bf k};{\bf p}_1;% 
{\bf p}_2 ) %
= && \langle \varphi_n \mid %  
\exp( i {\bf p}_1 \cdot {\bf r}_1 % 
+ i {\bf p}_2 \cdot {\bf r}_2 ) \times %
\nonumber \\ 
&& \left( \alpha_{l,1} % 
\exp( i {\bf k} \cdot {\bf r}_1 ) \right. % 
\nonumber \\ 
&& \left. + \alpha_{l,2} % 
\exp( i {\bf k} \cdot {\bf r}_2) \right) % 
\mid \varphi_0 \rangle,  
\label{e4} 
\end{eqnarray} 
where the indices $1$ and $2$ refer to
the first and second atomic electron,  
${\bf r}_j$ ($j=1,2$) are 
the coordinates of 
the $j$-th atomic electron 
in the frame $K_A$ 
and $\alpha_{l,(j)}$ are 
the Dirac matrices 
for this electron.   

The components of the inelastic form-factor
of the ion $ F_{\alpha }^I( m0; {\bf k} ) $ 
are given by 
\begin{eqnarray} 
F^{I}_{0}(m0;{\bf k}) &=& - 
\langle \chi_m \mid %
 \exp(i{\bf k} \cdot \mbox{\boldmath$\xi$})   %
\mid \chi_0 \rangle  %
\nonumber \\
F^{I}_{l}(m0;{\bf k}) &=& % 
\langle \chi_m \mid %
\alpha_{l} \, % 
 \exp(i{\bf k} \cdot \mbox{\boldmath$\xi$})   %
\mid \chi_0 \rangle,   %
\label{e5}
\end{eqnarray}
where $\mbox{\boldmath$\xi$}$ are 
the coordinates of the 
electron of the ion 
in the frame $K_I$. 

In the frame $K_A$ the momentum ${\bf q}^A$ 
transferred to the atom is given  by  
\begin{eqnarray}
{\bf q}^A &=& \left( {\bf Q}, q_{min}^A \right) 
\nonumber \\ 
q_{min}^A &=& \frac{ \varepsilon_n-\varepsilon_0 }{ v } + % 
\frac{ \epsilon_m-\epsilon_0 }{ \gamma v }, 
\label{e6}
\end{eqnarray}
where $q^A_{min}$ is the component of 
this momentum along the velocity ${\bf v}$ and 
represents the minimum momentum transfer 
to the atom in the frame $K_A$.  
The momentum transfer ${\bf q}^I$ 
to the ion, viewed in the frame $K_I$, 
reads 
\begin{eqnarray}
{\bf q}^I &=& (-{\bf Q}, -q_{min}^I) 
\nonumber \\ 
q_{min}^I &=&\frac{\epsilon_n-\epsilon_0}{v} + % 
\frac{\varepsilon_m-\varepsilon_0}{\gamma v}, 
\label{e7}
\end{eqnarray} 
where $q^I_{min}$ is 
the minimum absolute value of the 
momentum transfer to the ion 
in the frame $K_I$. Note that in the limit 
$c \to \infty$ the momenta (\ref{e6}) and (\ref{e7}) 
go over into their nonrelativistic counterparts. 

The eikonal model represents a good approximation 
provided the conditions $Z_I \stackrel{<}{\sim} v$ 
and $v \gg Z_A$ are fulfilled.  
At $\nu = Z_I/v \ll 1$ the amplitude 
(\ref{e1}) goes over into the first order 
transition amplitude \cite{VNU} but very 
substantially differs from the latter 
when $\nu $ approach $1$. In contrast to the 
first order amplitude, the amplitude (\ref{e1}) 
takes into account all six pair-wise interactions between 
the constituents of the projectile 
and the target \cite{f-add-1}. 

Below we shall consider only collisions in which 
the target gets ionized. 

In collisions at $Z_I \sim v \sim c$ 
the spectra of electrons emitted from the 
target and projectile show strong effects 
caused by both the relativity and the multiple 
exchanges of virtual photons between the colliding 
ion and atom. We, however, shall discuss 
these spectra elsewhere and here concentrate 
on the spectrum of the target recoil ions 
given as a function of the longitudinal component 
$p_{R,\parallel}$ of the recoil momentum 
($p_{R,\parallel} \parallel v$). 

\begin{figure}[t] 
\vspace{-0.35cm} 
\begin{center}
\includegraphics[width=0.31\textwidth]{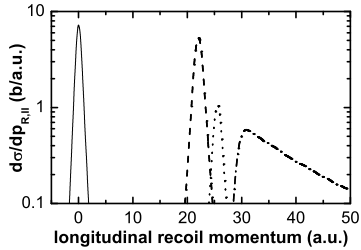}
\end{center}
\vspace{-0.9cm}
\caption{ \footnotesize{ The longitudinal 
momentum spectrum of the target recoil 
ions produced in  
$100$ MeV/u Nd$^{59+}$(1s)+H(1s) collisions.  
Dash, dot and dash-dot curves 
at large $p_{R,\parallel}$ correspond 
to the Nd$^{59+}$(n=2), Nd$^{59+}$(n=3) and 
Nd$^{60+}$ + e$^-$ states of the projectile, 
respectively. Besides, solid  
curve displays the total contributions 
of the above channels to the recoil spectrum 
at small $p_{R,\parallel}$. } } 
\label{figure1} 
\end{figure}

In figure \ref{figure1} results are shown for  
the longitudinal momentum spectrum 
of the target recoil ions, $d \sigma/d p_{R,\parallel}$,   
generated in the processes of the projectile electron 
loss and excitation to states with $n=2$ and 
$n=3$, where $n$ is the principal quantum number, 
in collisions of $100$ MeV/u Nd$^{59+}$(1s) 
with atomic hydrogen.  
The recoil peak centered at $p_{R,\parallel}=0$   
arises due to the two-center dielectronic interaction  
and thus is basically the effect of the first order 
in the projectile-target interaction. Therefore, 
it is predicted by both the first order 
and eikonal theories \cite{f-new}. 
The peaks at much larger $p_{R,\parallel}$, 
however, are not predicted by the first order theory 
and appear in the calculation only if 
all the interactions between the  
nuclei and the electrons are taken 
into account. Hence, these peaks represent  
clear signatures of the higher order effects 
in the interaction between the colliding particles. 

\begin{figure}[t] 
\vspace{-0.35cm}
\begin{center}
\includegraphics[width=0.5\textwidth ]{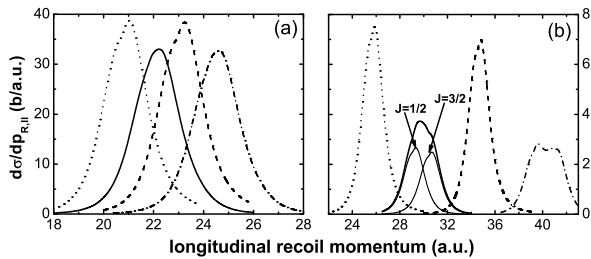}
\end{center}
\vspace{-1.1cm}
\caption{ \footnotesize{ The projectile 
excitation into states with the principal 
quantum number $n=2$ accompanied by target 
single ionization in collisions of  
$100$ MeV/u Nd$^{59+}$(1s) (a) and 
$325$ MeV/u U$^{91+}$(1s) (b) with He(1s$^2$).  
Dash and solid curves display results of the purely 
nonrelativistic ($c = \infty$) and fully relativistic 
calculations, respectively. 
Dot curves were obtained by 
assuming $c = \infty$ only in 
the description of the 
internal electron states. 
Dash-dot curves were calculated  
by setting $c = \infty$ only in the treatment of 
the relative ion-atom motion. } } 
\label{figure2}
\end{figure}

In figure \ref{figure2} we display  
results for the projectile excitation 
into states with $n=2$ accompanied 
by the single ionization of helium target.  
We see that the relativistic 
and nonrelativistic calculations 
predict rather different positions   
and shape of the target recoil peak.   

The relativistic effects influencing the form 
of the recoil spectrum can be subdivided into 
those depending on the collision velocity $v$ 
and those related to the inner motions of electrons 
within the colliding centers. 
As is seen in figure \ref{figure2}, 
in the momentum spectrum these two groups 
of the effects counteract. 

The relativistic effects 
in the inner motion of 
the electron in the ion lead 
to the energy splitting of 
the levels having the total 
angular momentum $j=1/2 $ and $j=3/2$ and 
increase the energy differences between 
them and the ground state.  
Because of the splitting 
the peak in the recoil momentum spectrum 
becomes broader and the splitting 
is also partly responsible 
for the decrease in the height of the peak. 
As a result of the increase in the energy differences   
the peak position is shifted to larger 
values of $p_{R,\parallel}$. However,  
the relativistic effects due to the relative ion-atom 
motion soften the recoil of the residual atomic core 
which shifts the peak position to lower values 
of $p_{R,\parallel}$. In the examples shown in figure 2 
the value of the collision velocity  
is very close to the value of the typical 
orbiting velocity of the electron in the 
ground state of the ion but 
the shift of the recoil peak due to 
the relativistic effects in the relative  
ion-atom motion turns out to be 
larger by about a factor of $2$.  
\begin{figure}[t] 
\vspace{-0.35cm}
\begin{center}
\includegraphics[width=0.31\textwidth]{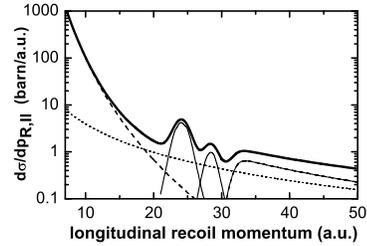}
\end{center}
\vspace{-0.9cm}
\caption{ \footnotesize{ The longitudinal  
momentum spectrum of He$^+$ recoils produced  
in $430$ MeV/u Th$^{89+}$(1s)+He collisions. 
Dash and dot curves show the contributions 
to the spectrum from the singly inelastic 
channel given by the interaction between 
the ionized electron and 
the residual atomic ion and by the 
nucleus-nucleus Rutherford scattering, respectively.  
Thin solid curves: the contribution from 
the projectile electron loss and  
excitations into $n=2,3$.  
Thick solid curve: the total contribution from 
the above channels. 
The contribution from 
the projectile excitation into bound states 
with $n\geq 4$ was not calculated.} } 
\label{figure3}
\end{figure}

The processes considered above   
are of course not the only ones 
which produce target recoil ions. 
Therefore, the important question 
is whether the signatures of  
the projectile-electron excitation and loss 
in the momentum spectrum of the target recoil ions
will not be masked by other processes 
(especially if the final state of the projectile 
is not detected). To answer this question 
we have performed a more extensive study 
of the target recoil momentum spectrum 
produced in collisions in which 
the initial internal state of the projectile 
may change or may remain the same.  
Some results of this study are shown in   
figure \ref{figure3} which actually 
illuminates the very general features 
in the formation of the longitudinal 
recoil spectrum in collisions 
with very heavy hydrogen-like ions 
at moderate values of $\gamma$.  
According to this figure 
one can separate four different regions 
of $p_{R,\parallel}$ in which 
the formation is dominated 
by qualitatively different mechanisms.  

(a) At small values of $p_{R,\parallel}$ 
the recoil spectrum is generated via  
the indirect coupling between the projectile, 
whose internal state {\it is unchanged},  
and the target core: the projectile induces 
a transition of the target electron and 
the electron exchanges its momentum 
with the target core.  
In relativistic collisions this channel 
is characterized by very large impact parameters 
($\gg 1$ a.u.) and is very efficient in transferring 
small values of the momentum to the target recoil. 
At larger $p_{R,\parallel}$, however, 
it rapidly loses its effectiveness 
which leads to the very strong decrease 
in the longitudinal spectrum when 
$p_{R,\parallel}$ growths. 
 
(b) Eventually with increasing $p_{R,\parallel}$ 
the direct coupling between the nuclei  
of the ion and atom (the Rutherford scattering) 
may start to dominate the formation 
of the spectrum. 
This channel 'works' in collisions 
with very small impact parameters   
($ \sim Z_I Z_A/v \sqrt{ M_A v p_{R,\parallel} } $,  
where $M_A$ is the mass of the atomic nucleus) 
and, provided $Z_I \sim v$, the nuclear scattering 
is accompanied by {\it target ionization}  
with the probability close to $1$ 
but the inner state of the projectile 
{\it remains unchanged} \cite{rutherford}.   

(c) With the further increase in $p_{R,\parallel}$ 
($3Z_I^2/8v \stackrel{<}{\sim} p_{R,\parallel} % 
\stackrel{<}{\sim} Z_I^2/v $)
the channel involving the excitation and 
loss of the tightly bound electron of the projectile 
comes into the play. Compared 
to the nucleus-nucleus collisions this channel 
is characterized by much larger values of the typical 
impact parameters, which at $Z_I \sim v$ are of the order of 
the size of the projectile ground state, 
and is much more effective. 
Since the absolute charge of the electron 
constitutes just about $1\%$ of the net ion charge 
and its mass is negligibly small compared to 
the mass of the nuclei,
such an effectiveness is quite surprising   
and may have interesting consequences.   

Consider, for instance, two very heavy projectiles   
whose atomic numbers differ just 
by $1$ and whose masses are very close 
(e.g. $^{209}_{\,\,\,83}$Bi ($A=208.98$) 
and $^{209}_{\,\,\,84}$Po ($A=208.98$), 
$A$ is the atomic mass). Let one of them be represented 
by a bare nucleus with a charge 
$Z_I^{(1)}$ and the second be a hydrogen-like 
ion with a net charge $Z_I^{(2)}-1 = Z_I^{(1)}$. 
Thus, the projectiles possess practically   
equal charge-to-mass ratios and seem to behave 
almost identically in external electric fields. 
However, if these projectiles 
collide with atoms, the atoms 
can easily 'recognize' 
whether the projectile 
is a bare nucleus or an hydrogen-like ion 
since in the latter case the spectrum 
of the target recoil ions possesses 
a prominent resonance-like structure, 
reflecting the excitation and loss of 
the projectile's electron (superimposed on the 
smooth background from collisions elastic 
for the projectile).  

(d) At even higher $p_{R,\parallel}$ the 
recoil spectrum is formed 
almost solely by the nucleus-nucleus 
scattering \cite{f2}.  

In very asymmetric collisions  
($M_I \gg \gamma M_A$, where $M_I$ is 
the mass of the ion nucleus)   
the contributions to the cross section  
$d\sigma/d p_{R,\parallel}$ due to 
the nuclear Rutherford scattering and 
due to the excitation of the electron of the ion
scale as $Z_A^2/M_A$ and $Z_A^2$, respectively. 
Therefore, the relative strength of the 
latter channel in the spectrum formation 
is weakest for collisions with atomic hydrogen. 

In summary, we have considered 
the doubly inelastic collisions of 
relativistic heavy hydrogen-like ions 
with the lightest atoms and 
shown that the physics of such collisions 
is strongly influenced by 
the higher order effects 
in the ion-atom interaction.  
We have discussed the manifestations of the 
relativistic effects caused by 
both the relative ion-atom motion 
and the electron motion in the internal 
states of the ion.   
Our results have enabled one to draw 
the general picture of the formation 
of the longitudinal momentum spectrum 
of the target recoil ions produced 
in high-energy collisions with 
very heavy hydrogen-like ions.   
These results also show 
a surprisingly strong effect of 
the projectile electron   
on the momentum transfer to the 
atomic recoil ion. 

In collisions of very highly charged ions 
with atoms the doubly inelastic 
cross sections are by several orders 
of magnitude smaller 
than the cross section 
for the pure atomic ionization.  
However, since at moderate $\gamma$ 
the regions of $p_{R,\parallel}$
relevant to the singly and doubly 
inelastic processes are well separated 
and because of the unexpectedly 'prominent'  
behavior of the projectile electron 
in the transfer of large $p_{R,\parallel}$, 
a great deal of information about  
the doubly inelastic collisions with 
relativistic heavy ions could be obtained 
in an experiment just by measuring the 
longitudinal momentum spectrum of 
the target recoil ions.


\begin{thebibliography}{99}
\bibitem{f-add-1} 
The doubly inelastic transition amplitude 
for collisions with atomic hydrogen   
can be obtained from Eq.(\ref{e1}) 
if the integration over, say, ${\bf p}_2$ 
is removed and the function $f(p_2,\nu)$ 
is set to $1$. Besides, in the rest of 
the integrand of (\ref{e1}) $p_2$ should 
be set to $0$ and the obvious changes have 
to be made in the atomic form-factors.
\bibitem{f-new} In collisions with highly charged 
ions the two-center dielectronic interaction 
is modulated by multiple interactions between 
the ionic nucleus and the electron of the target 
which, under the conditions of figure \ref{figure1}, 
decreases the height of the peak at 
$p_{R,\parallel} \simeq 0$ by about  
$25 \%$ compared to the prediction of 
the first order theory.
\bibitem{rutherford} To describe 
this channel of momentum transfer 
Eqs.(\ref{e6}) have to be corrected 
by adding a term $Q_{\perp}^2/(2M_A v)$ 
into the equation for $q_{min}^A$
(a similar change has to be made 
also in (\ref{e7})).  
Note that this correction becomes important  
only at very small impact parameters   
whose contribution to the doubly inelastic 
cross sections is negligible and,  
therefore, can safely be neglected when 
the latter are calculated.
\bibitem{f2} So far we have not mentioned 
the capture of a target electron by the projectile. 
This process occurs via the radiative 
and coulomb capture channels.   
At the impact energies under consideration  
the last channel is quite weak and 
is characterized by very small recoil peaks  
located at large negative $p_{R,\parallel}$ 
($p_{R,\parallel}= - (I_f/\gamma-I_i)/v % 
- m_e c^2(1-1/\gamma)/v $, where 
$I_i$ and $I_f$ are the initial and final 
binding energies of the electron). 
The relatively strong radiative capture channel   
results only in quite low values of the 
target recoil momenta where its effect 
is fully masked by the direct target ionization.
\end{thebibliography}
\end{document}